\documentclass[12pt]{article}
\usepackage[cp1251]{inputenc}
\usepackage{epsfig}

\begin{document}

\title{Comment on "Vector and axial-vector mesons at finite temperature".}
\author{M.Dey \\
Department of Physics, Presidency College,\\
Calcutta 700 073, India\\
\\
V.L.Eletsky and B.L.Ioffe\\
Institute of Theoretical and Experimental Physics,\\
B.Cheremushkinskaya 25, 117218 Moscow,Russia}
\date{}
\maketitle

\begin{abstract}
It is shown, that the correlators of vector and axial-vector currents at
finite temperature $T$ in order $T^2$ reduce to mixing of vacuum $VV$ and $AA
$ correlators with universal mixing coefficient given by the parameter $%
\varepsilon = T^2/6 F^2_{\pi}$, unlike the claim in the recent paper by
Mallik and Sarkar [1].
\end{abstract}

In a recent paper [1] Mallik and Sarkar considered vector and
axial-vector correlators at finite temperature in order $T^2$.
Earlier, the same problem was treated in our paper [2]. Mallik and
Sarkar criticize the calculation method used in [2] and claim,
that the results obtained in that paper are wrong. In this comment
we show that this is not true.

Let us remind shortly the basic results of [2]. \ \

The thermal correlators ($J = V,A$)

\begin{equation}
C_{\mu \nu }^{J}(q,T)=i\int ~d^{4}xe^{iqx}\sum\limits_{n}\langle n|T\{J_{\mu
}^{a}(0_{)},~J_{\nu }^{a}(x)\}exp[(\Omega -H)/T]|n\rangle
\end{equation}
were considered at low $T$ in order $T^{2}$. Here $e^{-\Omega
}=\sum\limits_{n}\langle n|exp(-H/T)|n\rangle $ and the sum is over the full
set of eigenstates of Hamiltonian $H$. The currents $V_{\mu }^{a}$ and $%
A_{\mu }^{a}$ are isovector vector and axial currents. The validity of the
chiral effective theory (CET) -- a low energy consequence of QCD -- was
taken as granted. The explicit form of the Hamiltonian was not specified. It
was assumed, that pions are massless and current algebra relations were
exploited. In the sum over $n$ in order $T^{2}$ it is sufficient to account
for the contribution of two lowest states -- vacuum and one-pion. Then the
factor $T^{2}$ arises simply from the one-pion phase space. The
contributions of higher excited states are suppressed as $T^{k}$,~ $k>2$ or
exponentially, $\sim e^{-m_{h}/T}$. It was found [2] that in order $T^{2}$
(at $q^{2}=-Q^{2},~Q^{2}\gg T^{2}$)

\begin{equation}
C_{\mu \nu }^{V}(q,T)=(1-\varepsilon )C_{\mu \nu }^{V}(q,0)+\varepsilon
C_{\mu \nu }^{A}(q,0)
\end{equation}
\begin{equation}
C_{\mu \nu }^{A}(q,T)=(1-\varepsilon )C_{\mu \nu }^{A}(q,0)+\varepsilon
C_{\mu \nu }^{V}(q,0),
\end{equation}
where $\varepsilon = T^2/6 F^2_{\pi}$, ~ $F_{\pi }=91MeV$ is the pion
decay constant. Therefore, in order $T^{2}$ the thermal $V$ and $A$
correlators reduce to their vacuum values, but in the vector correlator
there appears the admixture of the axial one and vice versa. The positions
of the poles (and other singularities) of thermal correlators are the same
as at $T=0$ (up to appearance of $A$ poles in $V$ channel and vice versa).
The residues of the poles are shifted in a universal way given by the
parameter $\varepsilon $. In ref.[3] more detailed arguments were presented
and the results were generalized to correlators of other currents (baryon,
etc.).

In their consideration of the same problem, Mallik and Sarkar
obtained the equation, equivalent to (2) with a difference, that
mixing coefficients are not universal and depend on the form of
low-energy Lagrangian. They assumed that the terms containing
$\rho $, $a_{1}$, $\omega $ and $f_{1}$ fields are added to the
standard CET Lagrangian. Then, in
their calculation of $C_{\mu \nu }^{V}(q,T)$ (2), besides the factor $%
1-\varepsilon $ in front of $C_{\mu \nu }^{V}(q,0)$ there appears a term
proportional to $g_{1}^{2}$, where $g_{1}$ is the $\rho w\pi $ coupling
constant. However, embedding of the $\rho w\pi $ interaction term into
effective low-energy Lagrangian is not legitimate. As can be easily proved
(using eq.(2.22) of [1]) this term violates $SU(2)\times SU(2)$ chiral
symmetry, leads to appearance of additionaal term in the expression for
axial current and, consequently, violates current algebra relations.
Moreover, famous low-energy theorems, which follow from current algebra
(Goldberger-Treiman, Adler, Gross-Lelwellyn-Smith, Weinberg sum rules etc.) 
would also be
violated if such a term persists. Therefore, the disagreement of the Mallik
and Sarkar results with eq.2 comes entirely from illegitimate form of the
low-energy Lagrangian used in [1]. If the $\rho w\pi $ interaction term in
the Lagrangian is omitted, then the results of [1] exactly coincide with
eq.2, found in [2].

\begin{figure}[tb]
\hspace{35mm} \epsfig{file=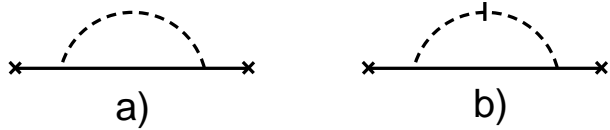, width=80mm}
\caption{Self-energy correction to correlator. Solid lines - mean vacuum
correlator, dashed lines -- pion propagator. a) Pion is virtual; b) pion is
from thermal bath.}
\end{figure}

Now, a few words about the criticizm of our paper [2], given in [1]. It was
claimed in [1], that the self-energy diagrams on Fig.1 (Fig.1b in [1],
Fig.1c,d in [2]) were ignored in [2]. In fact, there are two types of such
diagrams: 1) Fig.1a, where the pion is virtual. Such diagram corresponds to
vacuum correlator, is $T$-independent and is accounted in $C_{\mu \nu }(q,0)$
in (2),(3); 2) Fig.1b, where pion comes from thermal bath and is on the mass
shell. For a pion on mass shell its interaction vertex is proportional to
momentum (Adler theorem). Therefore, the contribution of the Fig.1b diagram
is $\sim T^{4}$ and may be omitted. (We repeat here the arguments of ref.[2]
for clarity).

To conclude: the results, no correlator mixing, obtained in [1] are valid only 
for the model Lagrangian
used in [1], but not for QCD and its consequense - low energy chiral effective
theory (CET). On the other hand our paper [2] based on QCD and CET shows correlator 
mixing at low temperature, The thermal $V$ and $A$ correlators mixing in order $T^2$ 
- eq. (2),(3) was confirmed in recent paper [4]. Hence the claim of the authors in [1]
that our paper [2] is "incorrect" is not correct.

\end{document}